\documentclass[twocolumn,showpacs,preprintnumbers,amsmath,amssymb,floatfix]{revtex4}
\usepackage{amsmath,amssymb}
\usepackage[dvips]{graphicx}
\usepackage{epsfig}
\usepackage{dcolumn}
\usepackage{bm}

\newcommand{\bs}{\begin{subequations}}
\newcommand{\es}{\end{subequations}}
\newcommand{\vp}{\varphi}
\newcommand{\JJ}{0-$\pi$JJ }

\begin{document}

\title[Critical relations in symmetric $0-\pi$ JJ]{Critical relations in symmetric $0-\pi$ Josephson junctions}

    \author{J.~A. Angelova$^*$ and T.~L. Boyadjiev$^\dag$}
\address{Dept. of Mathematics, University of Chemical Technology
and Metallurgy,
       Sofia 1756, Bulgaria}
\email{jordanka_aa@yahoo.com}
\address{Joint Institute for Nuclear Research, Dubna 141980, Russia}
\email{todorlb@jinr.ru}

\date{\today}

\begin{abstract}

Numerical modeling of dependences ``critical current -- external magnetic field'' for geometrically symmetric $0-\pi$ Josephson junctions is performed. The calculation of critical current is reduced to non-linear eigenvalue problem. The critical curve of the contact is obtained as an envelope of the bifurcation curves of different distributions of the magnetic flux.

The structure of vortices in contact is observed explicitly and the dependence of the basic physical characteristics of these vortices on junction's length is explored.

The comparison of numerical results and known experimental data shows good qualitative and quantitative conformity.
\end{abstract}


\maketitle

\section{Introduction}

The opportunity of existence and interest of studying of $0-\pi$ Josephson contacts (further on referred as \JJ s) have been shown in classical papers \cite{bul_77}, \cite{bul_78} (see the short history of the problem in \cite{bjzm_06}). For last years in the specified area significant number of theoretical and experimental works are published (see for example, \cite{gkk_02} -- \cite{sdg_07} and references therein).

In this article a numerical modeling of some experimental data \cite{wkgkwkk_06} concerning the relations ``critical current -- external magnetic field'' for such a junctions, is carried out.

For this purpose the transitions from Josephson to resistive regime by change the external current $\gamma$ are mathematically interpreted as a bifurcation of one of possible static distributions of the magnetic flux $\vp(x)$ existing in the junction for given external magnetic field $h_e$ (from now on all the variables are dimensionless).

Every solution of non-linear boundary value problem (BVP) for $\vp(x)$ generates some regular Sturm-Liouville problem (SLP). The minimal eigenvalue (EV) $\lambda_0$ of SLP allows to decide whether the concrete solution is stable or unstable in linear approximation. Note, that such approach for the traditional model of ``one-dimensional'' long josephson junctions (JJs) was proposed in \cite{ftbk_77, galfil_84}. 

In such approach the critical current $\gamma_c$ of some fixed distribution is that value of $\gamma$, for which the minimal EV $\lambda_0$ is vanishing at the given external magnetic field $h_e$. In order to calculate points of bifurcation $\gamma_c$ a non-linear eigenvalue problem is formulated, such that the current $\gamma$ is considered as an EV under the rest given junction's parameters including $\lambda_0$ \cite{bpp_jinr88}, \cite{bpp_nma88}. Since for given external magnetic field the static BVP can have more than one solution, then the critical current of JJ is determined as maximal of critical currents of possible stable distributions. Thus the critical curve (CC) of the junction can be constructed as an envelope of bifurcation curves (BCs) of different distributions of the magnetic flux.

\section{Statement of the problem}
Let consider $0-\pi$ junction of length $2l$, $l < \infty$, disposed along axis $x$, such that 0-junction (0JJ) is to the left and $\pi$-junction ($\pi$JJ) is to the right of the barrier $x = \zeta \in (-l,l)$. The junction is geometrically symmetric if $\zeta = 0$. In case of  overlap geometry the non-linear BVP for static distributions of magnetic flux $\vp(x)$ can be written as
\bs \label{bvp}
    \begin{gather}
        -\vp_{xx} + j_C(x) \sin \vp - \gamma =0, \label{bvpb}\\
        x \in (-l,\zeta)\cup (\zeta,l), \nonumber \\
        \vp (\zeta - \varepsilon ) = \vp (\zeta + \varepsilon), \label{cont1} \\
        \vp_x (\zeta - \varepsilon ) = \vp _x (\zeta + \varepsilon), \label{cont2} \\
        \vp_x(\pm l) - h_e =0, \label{bcl},
    \end{gather}
\es
where $j_C(x)$ is the amplitude of Josephson current
\[ j_C (x) = \left\{ {\begin{array}{*{20}c}
     1,\quad x \in [-l,\zeta]; \\
    -1,\quad x \in (\zeta,l],
 \end{array} } \right. \]

Possible solutions of \eqref{bvp} belong to a class $\mathbb{C}^1 [-l,l]$, i.e., continuously differentiable (smooth) functions. Physically this means that the magnetic field inside the junction is continuous only, and at the sewing point $x = \zeta$ continuity conditions  \eqref{cont1} and \eqref{cont2} are fulfilled \cite{gkk_02}-\cite{wkgkk_06}.

One can check that for $\gamma = 0$ the common solution of equations \eqref{bvpb} in the corresponding intervals is expressed by the elliptic functions, see \cite{os_67}. Under the given model's parameters $p = \{l, \zeta, h_e, \gamma\}$ the integration constants can be obtained by means of two boundary conditions \eqref{bcl} and two continuity conditions \eqref{cont1} and \eqref{cont2} \cite{galfil_84}.

For small magnitudes of the current $|\gamma| << 1$ the solution of \eqref{bvp} can be derived applying perturbation theory methods. But in all cases the solution of arising non-linear algebraic systems can be obtained only numerically. Such approach for deriving bifurcation curves, which correspond to variation of parameters $p$, even $\gamma  = 0$, meets serious difficulties. Therefore, in the present paper the direct numerical modeling of the problem \eqref{bvp} is preferred.

\section{Stability of static solutions}

Further we shall suppose continuous dependence of solutions
$\vp(x,p)$ of \eqref{bvp} on parameters $p \in \mathcal{P}$,
$\mathcal{P} \subset \mathbb{R}^4$. By now, the solutions
dependence on $p$ we shall mark only if it is necessary.

In order to analyze the stability of solutions of \eqref{bvp} under parameters variation we put in correspondence to each solution $\vp(x,p)$ a regular Sturm-Liouville problem
\bs  \label{slp}
    \begin{gather}
       -\psi_{xx} + q(x)\psi = \lambda \psi, \label{j1}\\
        x \in (-l,\zeta) \cup (\zeta,l), \nonumber \\
        \psi(\zeta-\varepsilon) - \psi(\zeta+\varepsilon) = 0, \\
        \psi_x(\zeta-\varepsilon) - \psi_x(\zeta+\varepsilon) = 0, \\
        \psi_x(\pm l) = 0, \\
        \int\limits_{-l}^l \psi^2(x)\,dx - 1 = 0, \label{norm}
    \end{gather}
\es
where the potential $q(x) = j_C(x) \cos\vp(x)$. Here $\lambda$ is a spectral parameter and \eqref{norm} represents the norm condition.

It is easy to prove the existence of non-degenerate discrete bounded below spectrum $\left\{\lambda_n\right\}_0^\infty$ of problem \eqref{slp}, see \cite{ls_88}. Then the bifurcation equation of order $n$ of some solution $\vp(x,p)$ is
\begin{equation}\label{bifn}
    \lambda_n(p) = 0, \quad n = 0, 1, \ldots
\end{equation}
It is convenient to interpret the last equality geometrically as a surface in the parameters' space $\mathcal{P}$. Each point on this surface is a bifurcation (critical) point of order $n$ for some solution of \eqref{bvp}. For fixed value of two of parameters, the equation \eqref{bifn} describes a curve on the plane defined by other two parameters --- a bifurcation curve (BC) of order $n$ of considered solution.

In this paper we mainly consider bifurcations of order 0 of solutions of \eqref{bvp} by change basic physical parameters --- the external magnetic field $h_e$ and the external current $\gamma$. An implicit equation of BC for some solution on $(h_e, \gamma)$ plane is
\begin{equation}\label{bif}
    \lambda_0\left(h_e, \gamma \right) = 0.
\end{equation}

For 0JJs bifurcations of static distributions of magnetic flux under variation of the length $l$ were studied, for example, in articles \cite{bt_sst02}, \cite{sbs_04}.

From mathematical point of view, the problem \eqref{bvp} can be
considered as a collection of necessary extremum conditions for
the total energy functional of the junction
\begin{equation}\label{fen}
    F[\vp ] = \int\limits_{ - l}^l
    {\left[ {\frac{1}{2}\vp_x^2 + j_C (x)
        \left( {1 - \cos \vp } \right) - \gamma \vp }
    \right]\,dx} - h_e \Delta \vp
\end{equation}
on a set of smooth on $[-l,l]$ functions $\vp(x)$.

Especially, the equations \eqref{bvpb} are Euler-Lagrange equations for \eqref{fen} at corresponding subintervals, and Weierstrass-Erdmann conditions yield \eqref{bcl}, \eqref{cont1}, and \eqref{cont2}, see \cite{gf63}. The total magnetic flux through the junction $\Delta\vp$ is defined traditionally
\begin{equation}\label{flux}
    \Delta\vp = \int\limits_{-l}^l \vp_x\,dx = \vp(l) - \vp(-l).
\end{equation}

SLP \eqref{slp} has to satisfy sufficient extremum conditions for \eqref{fen}, see \cite{gf63}. If $\lambda_0 > 0$ for some solution $\vp(x)$ of \eqref{bvp}, then for the same solution the second variation of $F[\vp ]$ is positive definite and the functional \eqref{fen} possesses minimum. Therefore the solution $\vp(x)$ is stable in linear approximation.

If $\lambda_0(p) = 0$ for some $p \in \mathcal{P}$ then the solution $\vp(x,p)$ has at $p$ a bifurcation of zero order, and the equations \eqref{j1} represents the Jacobi equations for the functional \eqref{fen}. For brevity we shall call solutions at any parameter's bifurcation point as B-solutions.

In this article we apply the method for numerical construction of
relations \eqref{bif} proposed in \cite{bpp_jinr88, bpp_nma88}
(see also review \cite{review_07}). At given  $\lambda$ both
equations \eqref{bvp} and \eqref{slp} are considered as unique
non-linear eigenvalue problem with spectral parameter $h_e$ or
$\gamma$. To find eigenvalues and eigenfunctions
$(\vp(x),\psi(x))$ we use an algorithm based on continuous
analogue to Newton method \cite{namn}. The discretization of
corresponding linearized BVPs is performed by means of
spline-collocation method. Such approach leads to difference
scheme with block three-diagonal matrix. The scheme's accuracy
assessment, determined by Runge's method on the sequence of
uniform meshes with steps $h$, $h/2$ and $h/4$, is $O(h^4)$.

\section{Discussion of numerical results}

\subsection{Non-bifurcation static solutions}

\begin{figure}
    \includegraphics[width=7cm]{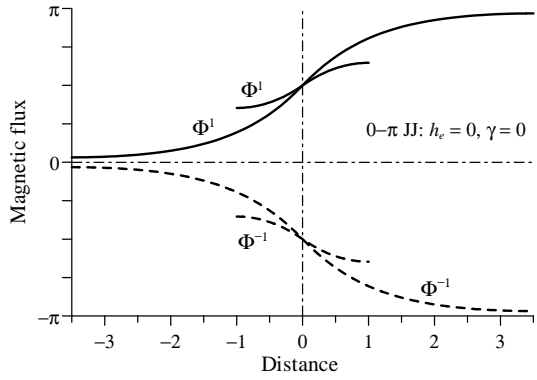}
    \caption{\label{F1a_pi_h0g0_f01} Magnetic flux in \JJ}
\end{figure}

First let's consider some solutions of \eqref{bvp}, which exist far enough from the bifurcation point of the magnetic field $h_e$ or/and current $\gamma$, and behavior of their main physical characteristics when parameters tend to the critical values.
\begin{figure}
    \includegraphics[width=7.3cm]{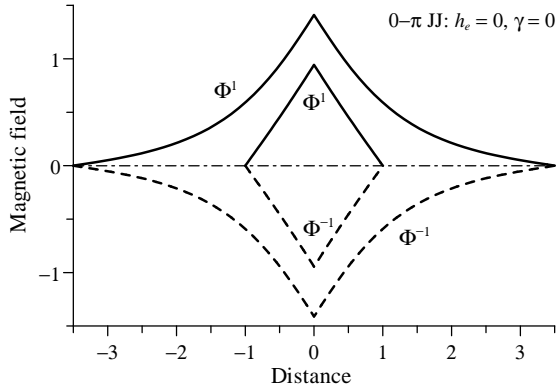}
    \caption{\label{S1a1_pi_h0g0_f2} Magnetic field in \JJ}
\end{figure}

On Fig.\ \ref{F1a_pi_h0g0_f01} the main stable states (of minimal
energy) of magnetic flux $\vp(x)$ for \JJ at two values of the
junction's length $2l = 2$ and $2l = 7$ in the field $h_e = 0$ and
under the current $\gamma = 0$ are demonstrated. Further, for
convenience, we shall use notation $\Phi^{\pm1}$ for mentioned
above solutions, where $\Phi^{1}$ is a fluxon and $\Phi^{-1}$ --
antifluxon.
\begin{figure}
    \includegraphics[width=6.4cm]{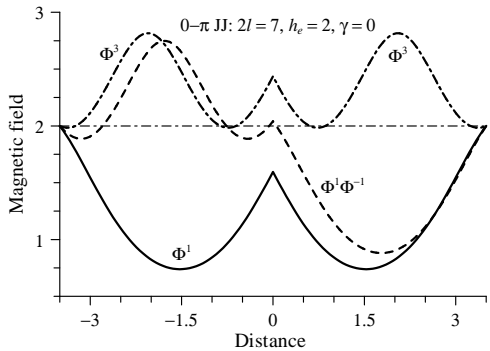}
    \caption{\label{Sts_pi_l7h2g0_f03} Bound states in \JJ at $h_e=2$}
\end{figure}

Corresponding distributions of magnetic field (derivatives
$\vp_x(x)$) are shown on Fig.\ \ref{S1a1_pi_h0g0_f2}. As in the
case of ring $\pi$-contact, see \cite{bul_77, bul_78}, the main
states have finite total magnetic field ($\Delta\vp \approx \pm
0.147$ at $2l=2$, and $\Delta\vp \approx \pm 0.468$ for $2l = 7$,
respectively), finite energy ($F[\Phi^{\pm 1}] \approx -0.04$, and
$F[\Phi^{\pm 1}] \approx -0.583$) and contain a half-integer
numbers $N[\Phi^{\pm 1}] = 0.5$ of fluxons (magnetic flux quanta),
where \cite{review_07}
\begin{equation}\label{nof}
    N[\vp] = \frac{1}{2l\pi} \int\limits_{-l}^{l} \vp(x)\,dx,
\end{equation}
is the average magnetic flux trough the junction.
\begin{figure}
    \includegraphics[width=6.4cm]{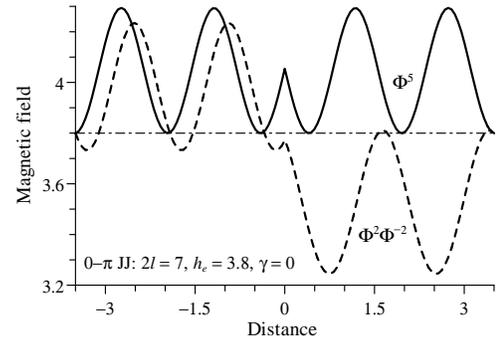}
    \caption{\label{S5s2_pi_l7g0_f04} Bound states in \JJ at $h_e=3.8$}
\end{figure}

At the center $x = 0$ we have $\Phi^{\pm 1}(0) = \pm \pi/2$ (see
Fig.\ \ref{F1a_pi_h0g0_f01}). Moreover, the conditions of
geometrical symmetry are fulfilled $\Phi^{-1}(x) = -\Phi^{1}(x)$,
$\Phi^{-1}_x(x) = -\Phi^{1}_x(x)$, and $\Phi^{\pm 1}_x(x) =
\Phi^{\pm 1}_x(-x)$.

\begin{figure}
    \includegraphics[width=6.4cm]{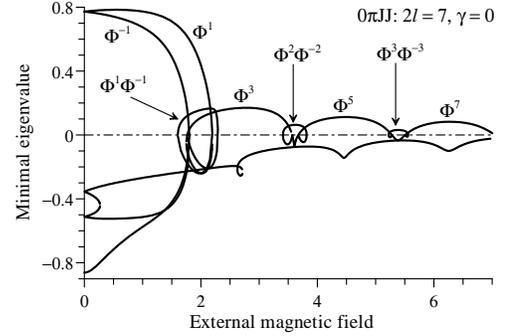}
    \caption{\label{Lah_pi_l7g0_f05} Dependence $\lambda_0(h_e)$ for \JJ\ of length $l=7$ at $\gamma = 0$}
\end{figure}

Except vortices demonstrated on Fig.\ \ref{F1a_pi_h0g0_f01}, at
$h_e = 0$ and $\gamma = 0$ there exist Meissner's states $M_0(x) =
0$ and $M_{\pi}(x) = \pi$ ($+ 2\pi k$, $k = 0, \pm 1, \ldots$).
Such distributions have zero total magnetic flux $\Delta\vp$  and
zero energy $F[\vp]$, but $N[M_0]=0$ and $N[M_\pi] = 1$. Note,
that the potential of SLP induced by $M_0$, is $q(x) = j_C(x)$ and
$q(x) = -j_C(x)$ for $M_\pi$. Hence no one of Meissner's solutions
can be stable in contrast to traditional 0JJ and $\pi$JJ, where
$q(x)=1$ for $M_0$ (stable) and $q(x) = -1$ for $M_\pi$
(unstable). In particular $\lambda_0 \approx -0.3$ at $2l = 2$ and
$\lambda_0 \approx -0.86$ for $2l = 7$. Numerical experiment
confirms, that Meissner's solutions remain unstable in all
admissible range of magnitudes $h_e$ and $\gamma$. Thereby in case
of \JJ the traditional Meissner's branches are missed on CCs, as
it is observed in the experiment \cite{wkgkwkk_06}.

\begin{figure}
    \includegraphics[width=6.4cm]{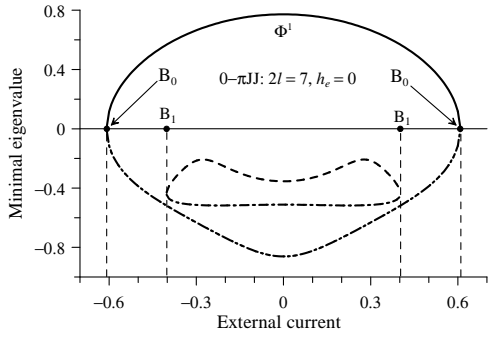}
    \caption{\label{Lagam_h0l7p_f06} Dependence $\lambda_0(\gamma)$ for \JJ\ of length $l=7$ at $h_e = 0$}
\end{figure}

Increasing value of $h_e$ arise new more complicated bound states
of the magnetic flux in the junction --- pure and mixed chains of
fluxons and antifluxons. Pure chains of vortices are ``composed''
only of fluxons or antifluxons. Mixed chains are obtained after
non-linear interaction between fluxons and antifluxons. To mixed
solutions of type $\Phi^{m}\Phi^{-m}$ correspond symmetric
$\Phi^{-m}\Phi^m$, $m = 1,2,\ldots$. On Fig.\
\ref{Sts_pi_l7h2g0_f03} chains of vortices  of internal magnetic
field $\vp_x(x)$ for the field $h_e=0$ and at the current
$\gamma=0$ are demonstrated: one-soliton distribution $\Phi^1$,
system of three non-linearly interacting solitons ($N[\Phi^3] =
2.5$), and also $\Phi^1\Phi^{-1}$, pair soliton -- antisoliton,
with $N[\Phi^1\Phi^{-1}] \approx 0.005$; the symmetrical pair
$\Phi^{-1}\Phi^{1}$ yields $N[\Phi^{-1}\Phi^{1}] \approx 0.995$.
Similarly, on Fig.\ \ref{S5s2_pi_l7g0_f04} a pure chain of fifth
solitons $\Phi^5$ ($N[\Phi^5] = 4.5$), and mixed chain
$\Phi^{2}\Phi^{-2}$ of two solitons and two antisolitons
($N[\Phi^{2}\Phi^{-2}] \approx 4.002$ at $h_e = 3.6$), pinning at
the inhomogeneity at the point $\zeta = 0$ are shown. For the
symmetrical pair $\Phi^{-2}\Phi^{2}$ we have $N[\Phi^{-2}\Phi^{2}]
\approx 2.998$.
\begin{figure}
    \includegraphics[width=7.5cm]{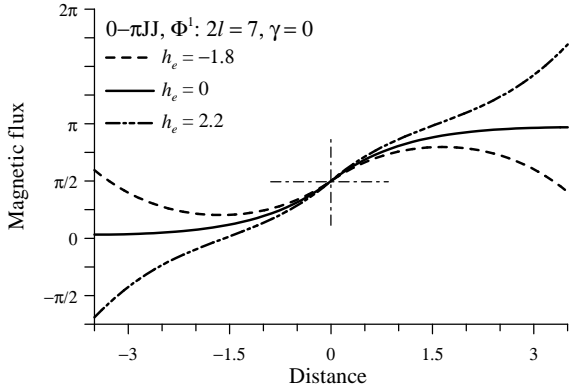}
    \caption{\label{F1_h_l7p_f07} Magnetic flux distributions $\vp(x)$ at $\gamma = 0$}
\end{figure}

On Fig.\ \ref{Lah_pi_l7g0_f05} the dependence of minimal EV of
\eqref{slp} on the external magnetic field $h_e$ for the junction
of length $2l = 7$ at $\gamma = 0$ is represented. In accordance
with \eqref{bif} the zeroes of curves $\lambda_0(h_e, 0)$ are
bifurcation points of the corresponding solutions. Each curve
possesses two zeros which correspond to the lower $h_{min}$ and
upper $h_{max}$ critical magnetic field for concrete distribution
$\vp(x)$. The distance $\Delta h = h_{max} - h_{min}$ represents
the domain of existence of the distribution $\vp(x)$ on the field
$h_e$ at $\gamma=0$. In particular, for $\Phi^1$-distribution
calculations yield $h_{min} \approx -1.8$ and $h_{max} \approx
2.2$, thus $\Delta h \approx 4$.
\begin{figure}
    \includegraphics[width=7.0cm]{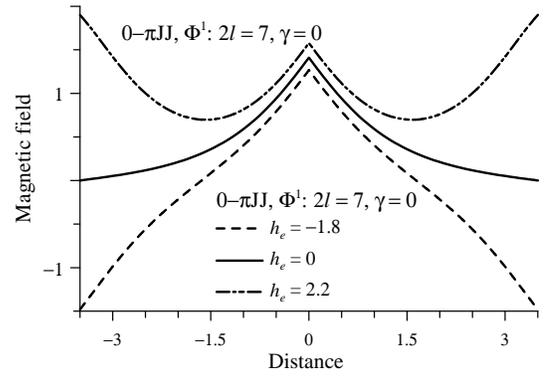}
    \caption{\label{S1_h_l7p_f_08} Magnetic field distributions $\vp_x(x)$ at $\gamma = 0$}
\end{figure}

Similarly, on Fig.\ \ref{Lagam_h0l7p_f06} the behavior of the
minimal EV of SLP by change the current $\gamma$ for $2l = 7$ and
$h_e = 0$ is demonstrated. Each upper-down curve corresponds to
two symmetrical vortex distributions with the same $\lambda_0$ and
same energy $F$,  but with opposite signs of number of fluxons and
$\Delta\vp$. The curve in the upper half-plane corresponds to
single fluxon distributions  $\Phi^1$ and $\Phi^{-1}$. Points
$B_0$ ($|\gamma| \approx 0.6$) are points of bifurcation of
mentioned solutions varying parameter $\gamma$. At these points
arises confluence of $\Phi^1$ with Meissner's distribution $M_0$,
which is much deformed  by the current. For the given junction,
the distance $\Delta\gamma \approx 1.21$ between points $B_0$
defines domain of stability of distributions $\Phi^1$ via current
$\gamma$ in the field $h_e=0$.
\begin{figure}
    \includegraphics[width=7.3 cm]{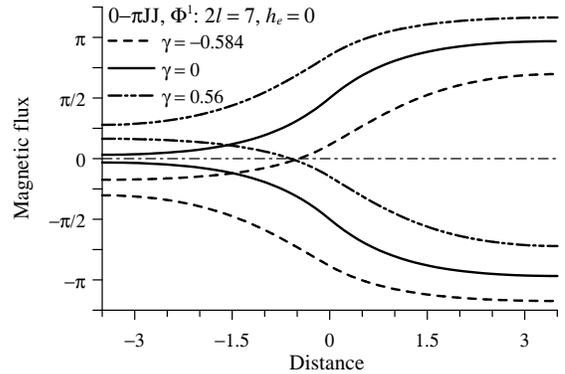}
    \caption{\label{F1_g_l7p_f09} Magnetic flux distributions $\vp(x)$ at $h_e = 0$}
\end{figure}

At points $B_1$ the first EV $\lambda_1$ of SLP is vanishing. This
means that there is a bifurcation of unstable distribution to
another unstable one. In the experiments it is difficult to
discover such transitions because of small life time
\cite{galfil_84}.

Fig.\ \ref{F1_h_l7p_f07} illustrated the deformation of the basic
fluxon $\Phi^1$ by external field in \JJ of length $2l = 7$ at the
current $\gamma = 0$. Corresponding distributions of magnetic
field $\vp_x(x)$ in contact are given at Fig.\
\ref{S1_h_l7p_f_08}. It is well observed, that the main changes
are localized at the neighborhood of contact's ends. The center
$\vp(0)$ of fluxon is conserved.

Similarly, on Fig.\ \ref{F1_g_l7p_f09} the deformation of fluxon
$\Phi^1$ by the current $\gamma$ at zero field $h_e$ is
represented. Positive current $\gamma$ displaces the fluxon's
graph upwards (the maximum of corresponding soliton $\vp_x(x)$
moves to the left), and the negative vice versa --- the graph of
$\vp(x)$ moves down, and the graph of soliton $\vp_x(x)$ --- to
the right.

The basic numerical characteristics of the first several
non-bifurcation magnetic flux vortices in contact at $2l = 7$ and
$\gamma = 0$ are given in the Table \ref{tabl1}.
\begin{table}[ht]
        \begin{tabular}{|c|c|c|c|c|c|c|c|}
        \hline
        Type & $h_e$ & $\lambda_0$ & $N[\vp]$ & $\Delta\vp/2\pi$&$\vp(0)/\pi$ & $F[\vp]/8$ \\
        \hline
        $\Phi^{-1}$         & 0    & 0.772  & $-0.5$ & $-0.468$ & $-0.5$ & $-0.586$  \\
        $\Phi^1$            & 0    & 0.772  & 0.5    &  0.468   & 0.5    & $-0.586$  \\
        $\Phi^{-1}\Phi^{1}$ & 1.61 & 0.0275 & 0.885  & 1.7      & 0.664  & $-1.349$  \\
        $\Phi^{1}\Phi^{-1}$ & 1.61 & 0.0267 & 2.115  & 1.7      & 2.336  & $-1.346$  \\
        $\Phi^{3}$          & 2    & 0.108  & 2.5    & 2.582    & 2.5    & $-1.883$  \\
        $\Phi^{-2}\Phi^{2}$ & 3.41 & 0.015  & 2.897  & 3.773    & 2.771  & $-5.144$  \\
        $\Phi^{2}\Phi^{-2}$ & 3.41 & 0.012  & 4.103  & 3.773    & 4.229  & $-5.140$  \\
        $\Phi^{5}$          & 4.6  & 0.111  & 4.5    & 5.049    & 4.5    & $-9.378$ \\
        $\Phi^{-3}\Phi^{3}$ & 5.28 & 0.029  & 4.963  & 5.882    & 4.866  & $-12.22$ \\
        $\Phi^{3}\Phi^{-3}$ & 5.28 & 0.023  & 6.037  & 5.882    & 6.134  & $-12.217$ \\
        $\Phi^{7}$          & 6.4  & 0.081  & 6.5    & 7.072    & 6.5    & $-18.00$ \\
        \hline
    \end{tabular}
\caption{Magnetic flux vortices in \JJ \label{tabl1} }
\end{table}

One can see that the pure magnetic flux distributions match to half-integer values of number of fluxons \eqref{nof}
$$N[\Phi^n] = n \mp \frac{1}{2},$$
where the minus sign correspond to $n>0$ and plus --- to $n < 0$, while for mixed fluxon states we have
$$N[\Phi^n\Phi^{-n}] + N[\Phi^{-n}\Phi^{n}] = 2n \mp \frac{1}{2}.$$

Let's note that the quotients $\vp(0)/2\pi$ fulfil the similar relationships as well.

\subsection{Bifurcations of static solutions}

In this article we shall mainly study bifurcations of static solutions varying external field $h_e$ and external current $\gamma$.

\begin{figure}
    \includegraphics[width=7 cm]{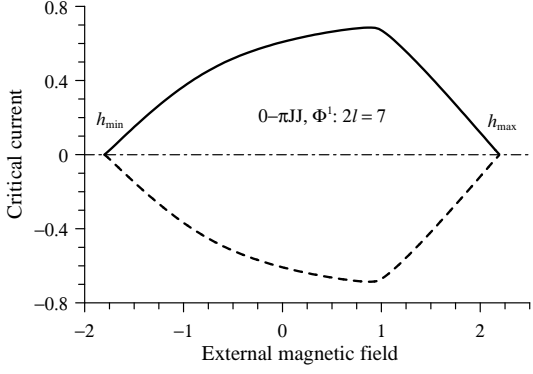}\caption{\label{Gcrh_f1_l7p_f10} BCs of the main fluxon $\Phi^1$}
\end{figure}
\begin{figure}
    \includegraphics[width=7 cm]{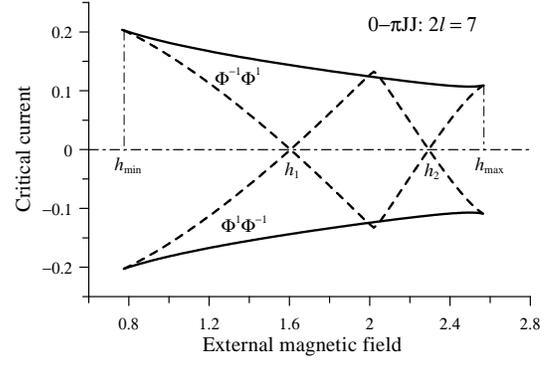}
    \caption{\label{Gcrh_f1a1_l7p11} BCs of pairs $\Phi^{-1}\Phi^1$, $\Phi^{1}\Phi^{-1}$}
\end{figure}
The locus on the plane $\left(h_e,\gamma\right)$ satisfying
equation \eqref{bif}, we shall call a bifurcation curve. A unique
BC corresponds to each solution of non-linear BVP \eqref{bvp}.
\begin{figure}
    \includegraphics[width=7 cm]{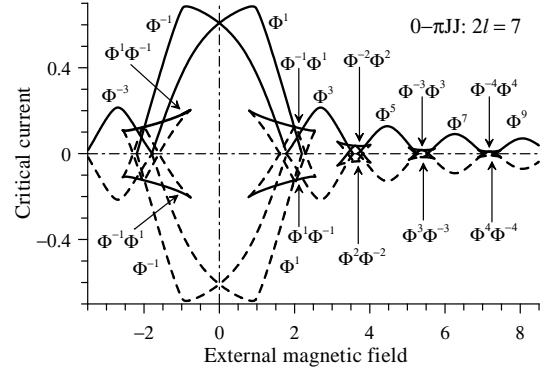}
    \caption{\label{Gcrh_l7p_f12} BCs of distributions in \JJ contact of length $2l=7$}
\end{figure}

For example, on Fig.\ \ref{Gcrh_f1_l7p_f10} the BC of the vortex
$\Phi^1$ in \JJ of length $2l = 7$ at $\gamma>0$ (solid line) and
$\gamma<0$ (dashed line) is demonstrated. The distance $\Delta h_e
\approx 4$ between zeroes $h_{min} \approx -1.8$ and $h_{max}
\approx 2.2$ represents the region of existence of $\Phi^1$ by
change the external magnetic field $h_e$. The graphs of
derivatives $\vp_x(x)$ of solutions in a small neighborhood of
bifurcation points $h_{min}$ and $h_{max}$ are shown on Fig.\
\ref{S1_h_l7p_f_08} by dotted and dash-dotted lines
correspondingly.
\begin{figure}
    \includegraphics[width=7 cm]{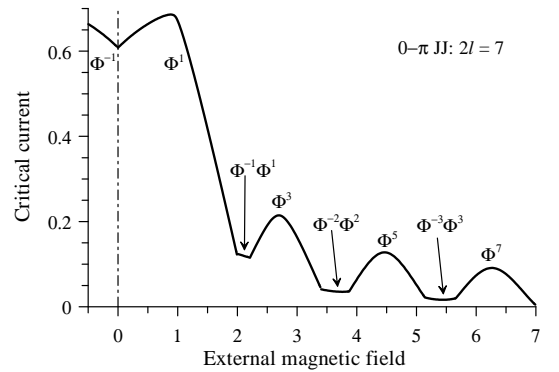}
    \caption{\label{Crc_l7op_f13} CC for \JJ\ of length $l=7$}
\end{figure}

Similarly, on Fig.\ \ref{Gcrh_f1a1_l7p11} the BCs for the pair
$\Phi^{-1}\Phi^1$, $\Phi^{1}\Phi^{-1}$ are demonstrated. The
existence domain of distributions via field $\Delta h = h_{max}-
h_{min}$ is separated by intermediate points $h_1$ and $h_2$, for
which $\gamma_{cr} = 0$, on three subdomains.
\begin{figure}[ht]
    \includegraphics[width=7 cm]{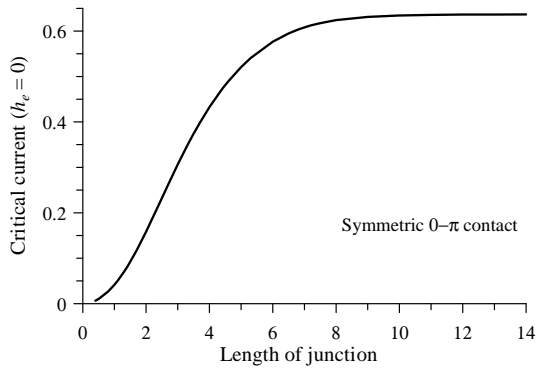}
    \caption{\label{Gcrl_h0pi_f14} Disposition of the point $\lambda_0(0,\gamma)$ as a function of length $l$}
\end{figure}

Let us note, that for $h_e \in \left(h_{min}, h_1\right)$ and $h_e
\in \left(h_2, h_{max} \right)$ the corresponding domains of
existence of distributions via current $\gamma$ are lied
over/under the horizontal axes $\gamma_{cr} = 0$.
Thus when $|\gamma|$ increases from zero, the mixed vortices at the mentioned domains have a current of ``birth'' (for given $h_e$ these are points on dotted curves) and a current of ``destruction'' --- points on solid curves. For $h_e = h_{min}$ and $h_e = h_{max}$ the currents of ``birth'' and ``destruction'' coincide.

At the points $h_e = h_1$ and $h_e = h_2$ the critical current of distributions
$\Phi^{-1}\Phi^1$ and $\Phi^{1}\Phi^{-1}$ is equal to zero.

The effect under consideration presents also in JJ with micro-inhomogeneities of resistive or shunt type \cite{review_07}.  Maybe, this effect can be determined experimentally using methods proposed in \cite{vdks_88}.
\begin{figure}[ht]
    \includegraphics[width=7.0 cm]{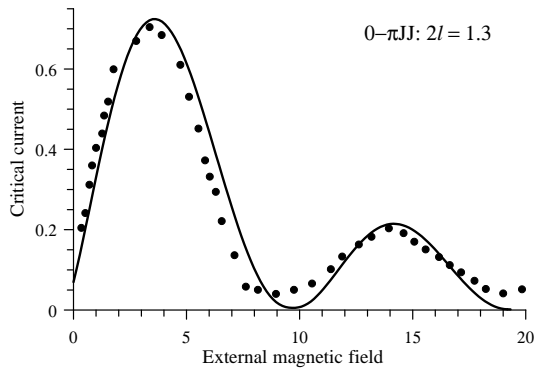}
    \caption{\label{Gcrh_l13p_f15} CC for \JJ of length $l = 1.3$}
\end{figure}

All possible BCs of distributions of magnetic flux in  \JJ of
length $2l = 7$, existing at widely differing $h_e$, are given on
Fig.\ \ref{Gcrh_l7p_f12}. The critical curve of the junction
contains all points of BCs with maximal modulo critical current at
a given field $h_e$, i.e., the contact's CC is derived as envelope
of BCs of all distributions. On Fig.\ \ref{Crc_l7op_f13} CC of the
contact mentioned above is illustrated for $h_e \le 10$. Because
of the symmetry, only the right upper quarter of CC is
demonstrated. Let us note, that CC in symmetric \JJ represents a
set of branches, which sequentially correspond to pure and
symmetric mixed distributions.

An important peculiarity is the shift of the critical current maximum $\gamma_{cr}$ to the right for $h_e>0$ (to the left for $h_e<0$) from the vertical line $h_e = 0$. The point $\gamma_{0} \equiv \gamma_{cr} (h_e=0)$ essentially depends on the contact's parameters --- (half)length $l$ and location $\zeta$ of the barrier. 

For the geometrically symmetric junction ($\zeta = 0$) the
dependence $\gamma_0(l)$ is given on Fig.\ \ref{Gcrl_h0pi_f14}.
For small $l$ the current $\gamma_0$ non-linearly quickly enough
tends to zero. At large length $l$ the current $\gamma_0$
asymptotically diverges to the constant value $\gamma_\infty
\approx 0.637$, while the total magnetic flux $\Delta\vp(2l) \to
\pi$.
\begin{figure} [ht]
    \includegraphics[width=7 cm]{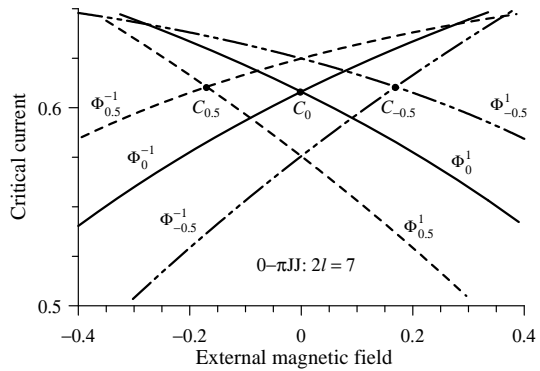}
    \caption{\label{Comp_l7pp_f16} Bias of intersection points of $\Phi^1$ \& $\Phi^{-1}$
    critical curves for different $\zeta$}
\end{figure}

A comparison of results of numerical modeling of CC (entire curve)
with available experimental data \cite{wkgkwkk_06} for short \JJ
of length $2l = 1.3$ is demonstrated on Fig.\ \ref{Gcrh_l13p_f15}.
Note, that for such length the contribution of BC of mixed states
on junction's CC can be neglected.

A good qualitative correspondence between numerical and existing
physical experiments is demonstrated. A plausible reason for some
quantitative differences can be the existence of different
in-symmetries in experimental samples, as $\zeta \ne 0$ and etc.
For example, at Fig.\ \ref{Comp_l7pp_f16} the influence of the
shift $\zeta \ne 0$ on $\Phi^{\pm 1}$ BCs in case $2l = 7$ is
demonstrated. For $\zeta = 0$ the intersection point $C_0$
correspond to $h_e|_{\zeta = 0} = 0$, while the inhomogeneity bias
to the left or right ($\zeta = \mp 0.5$) leads to bias of
corresponding intersection points $C_{\mp 0.5}$ to the right or
left ($h_e|_{\zeta = \mp 0.5} \approx \pm 0.17$) correspondingly.

\section{Conclusions}

We have numerically modelled experimental data \cite{wkgkwkk_06} for symmetric $0-\pi$ Josephson junctions.

Essentially, some stable bound states of magnetic fluxes and corresponding magnetic fields distributions are derived and demonstrated. For \JJ of length $2l=7$ we obtain and illustrate relations of minimal eigenvalue $\lambda_0$ on external magnetic field $h_e$ and external current $\gamma$. We illustrate numerically the important role of average magnetic field in the JJ as a measure of fluxon's number in a vortex chains. Special BCs and CCs for some fluxons and antifluxons are represented, too.

For relatively short \JJ a good qualitative conformity between
numerical and existing physical results takes place. We propose
that some quantitative differences can be caused on different
in-symmetries in experimental samples.

\begin{acknowledgments}
Authors thank Dr. Edward Goldobin for the statement of the
problem.
\end{acknowledgments}


\end{document}